\documentclass[twocolumn,showpacs,preprintnumbers,amsmath,amssymb]{revtex4}
\usepackage{graphicx}
\usepackage{dcolumn} 
\usepackage{bm} 
\usepackage{longtable}

\begin{document}

\title{Unconventional Fermi surface spin textures in the 
Bi$^{\ }_{\mathsf{x}\vphantom{1}}$Pb$^{\ }_{1-\mathsf{x}}$/Ag(111) 
surface alloy}
\author{Fabian Meier$^{1,2}$}
\author{Vladimir Petrov$^{3}$}
\author{Sebastian Guerrero$^{4}$}
\author{Christopher Mudry$^{4}$}
\author{Luc Patthey$^{2}$}
\author{J\"urg Osterwalder$^{1}$}
\author{J. Hugo Dil$^{1,2}$}
\affiliation{
$^{1}$Physik-Institut, Universit\"at Z\"urich, Winterthurerstrasse 190, 
CH-8057 Z\"urich, Switzerland 
\\ 
$^{2}$ Swiss Light Source, Paul Scherrer Institut, CH-5232 Villigen, 
Switzerland 
\\ 
$^{3}$St. Petersburg Polytechnical University, 29 Polytechnicheskaya St, 195251 St Petersburg, Russia
\\
$^{4}$ Condensed matter theory group,
Paul Scherrer Institut, CH-5232 Villigen, Switzerland
            }
\date{\today}

\begin{abstract}
{The Fermi and Rashba  
energies of surface states
in the Bi$^{\ }_{\mathsf{x}\vphantom{1}}$Pb$^{\ }_{1-\mathsf{x}}$/Ag(111) alloy
can be tuned simultaneously by changing the composition parameter 
$\mathsf{x}$.}
We report on unconventional Fermi surface spin textures 
observed by spin and angle-resolved photoemission spectroscopy
{that are correlated with 
a topological transition of the Fermi surface occurring at 
$\mathsf{x}=0.5$.}
We show that the surface state{s} 
remain fully spin polarized upon alloying 
and that the spin polarization vectors are 
approximately tangential to the constant energy contours. 
We discuss the implications of the {topological 
transition for the transport of spin}.
\end{abstract}


\maketitle

Controlling the spin degree of freedom of the electron lies at the heart 
of spintronics \cite{wolf}. One possibility to manipulate the electron spin 
without the need of any external magnetic field is found in the Rashba-Bychkov 
(RB) effect \cite{rashba}{.} 
It appears in (quasi) two-dimensional electron {or} hole 
systems with a lack of inversion symmetry and plays a prominent role 
for a proposed spin field-effect transistor
\cite{datta}{.} 
For most systems, the RB effect is small. Therefore, many of the related intriguing effects, 
such as a renormalization of the Fermi liquid parameters \cite{saraga}, 
changes in the electron-phonon coupling \cite{cappelluti}, 
enhanced superconductivity transition temperatures \cite{cappelluti2}, 
and real space spin accumulation \cite{sinova, wunderlich, Liu} 
remain for the most part experimentally unobservable.

\begin{figure}[t]
\begin{center}
\includegraphics[width=0.48\textwidth]{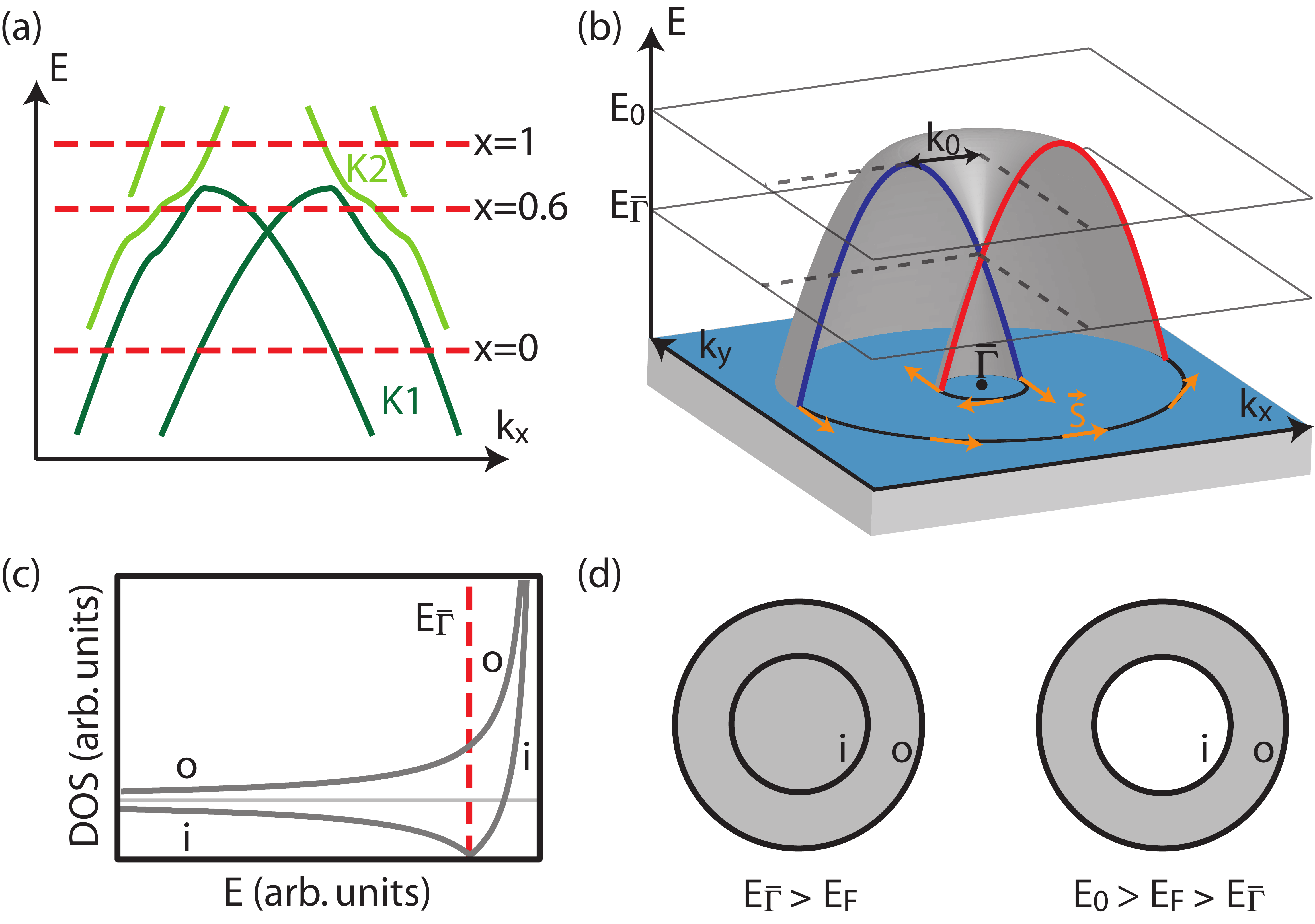}
\caption{
(color online) 
(a)
 {Qualitative plot} of the surface state band structure of 
Bi/Ag(111) ($\mathsf{x}=1$) 
{along the direction} $\bar{\Gamma}\bar{K}$ 
{in momentum space} 
(adapted from Ref.~\cite{bihlmayer})
showing the two Kramer's pairs K1 and K2.
As $\mathsf{x}$ is decreased, 
the Fermi level (dashed lines) lowers continuously 
and the spin splitting becomes smaller (not shown). 
(b)
Schematic picture of the Rashba effect for a two-dimensional hole gas around $\bar{\Gamma}$
and illustration of the relevant parameters.
The yellow (light gray) arrows are the spin expectation values of the eigenspinors.
(c) Density of states
for the outer 
($\mathrm{o}$) 
and inner 
($\mathrm{i}$) 
constant energy contours{.} 
(d) Hole Fermi 
{seas (gray regions) and Fermi surfaces (thick lines)} 
when
$E^{\ }_{\bar\Gamma}>E^{\ }_{F}$ 
and 
$E^{\ }_{0}>E^{\ }_{F}>E^{\ }_{\bar\Gamma}$.
        }
\label{Fig1}
\end{center}
\end{figure}

Recently{,} 
it has been shown that the {RB effect} 
is dramatically enhanced in the 
Bi/Ag(111)$(\sqrt{3}\times\sqrt{3})R30^{\circ}$ 
and 
Pb/Ag(111)$(\sqrt{3}\times\sqrt{3})R30^{\circ}$ 
surface alloys due to an additional in-plane inversion asymmetry 
\cite{ast,pacile,bihlmayer,premper,me}. 
Furthermore, the band structure can be continuously tuned between 
these two systems by {substituting Bi with Pb} 
\cite{astmixed}, 
as schematically illustrated in Fig.~\ref{Fig1}(a). 
The large {RB effect} combined with the tunability of the Fermi
and the {RB} energ{ies} make
Bi$^{\ }_{\mathsf{x}\vphantom{1}}$Pb$^{\ }_{1-\mathsf{x}}$/Ag(111)$%
(\sqrt{3}\times\sqrt{3})R30^{\circ}$, 
henceforth 
Bi$^{\ }_{\mathsf{x}\vphantom{1}}$Pb$^{\ }_{1-\mathsf{x}}$/Ag(111), 
an ideal model {RB system} 
to study the {geometrical and the topological
changes in the Fermi surface of its surface states \cite{cappelluti,aststm}.}
{It is clear that
the large conductivity of the Ag substrate short-circuits 
possible spin currents at the surface of
Bi$^{\ }_{\mathsf{x}\vphantom{1}}$Pb$^{\ }_{1-\mathsf{x}}$/Ag(111),
but {RB} semiconductors~\cite{Nitta} or thin metallic films~\cite{Dil} might be found that are equally tunable and
suited for technological applications.}

{
We present in this work spin and angle resolved photoemission spectroscopy (SARPES) data 
on surface states of 
Bi$^{\ }_{\mathsf{x}\vphantom{1}}$Pb$^{\ }_{1-\mathsf{x}}$/Ag(111) 
to resolve the changes in their Fermi surface spin textures 
(FSST) as a function of composition $\mathsf{x}$. 
We will argue that the spin transport
is strongly affected by a topological transition of the
Fermi surface taking place at the critical value 
{$\mathsf{x}^{\ }_{\mathrm{c}}=0.5$}.}

The {RB effect} occurs at interfaces or surfaces 
{whenever} the absence of the space inversion symmetry 
lifts the spin degeneracy due to the spin-orbit coupling. 
The simplest example of a {RB} Hamiltonian is given by  
\cite{winkler},
\begin{subequations}
\label{eq: def H RB}
\begin{eqnarray}
H=H^{\ }_{0}+H^{\ }_{\mathrm{RB}},
\label{H}
\end{eqnarray}
where the kinetic energy of the two-dimensional 
{hole gas with negative effective mass $m^{*}$} 
is 
\begin{eqnarray}
H^{\ }_{0}=
\sigma^{\ }_{0}
\left(
E^{\ }_{\bar\Gamma} -
\frac{\hbar^{2}}{2m^{*}}\boldsymbol{\nabla}^{2}
\right),
\label{eq: def H0}
\end{eqnarray}
while the {RB} term is
\begin{eqnarray}
H^{\ }_{{\mathrm{RB}}}=
-
\alpha^{\ }_{{\mathrm{RB}}}
\left(
i\sigma^{\ }_{y}\frac{\partial}{\partial x}
-
i\sigma^{\ }_{x}\frac{\partial}{\partial y}
\right).
\label{eq: def Hr}
\end{eqnarray}
\end{subequations}
The positive coupling constant
$\alpha^{\ }_{{\mathrm{RB}}}$ 
reflects the {RB coupling}. 
The unit $2\times2$ matrix is denoted by $\sigma^{\ }_{0}$, 
while $\sigma^{\ }_{x}$ and $\sigma^{\ }_{y}$ 
are the standard Pauli matrices in the basis in which the quantization axis 
is along the $z$ direction. 
The eigenenergies of $H$ yield {the upper 
$(+)$ and lower $(-)$ {RB} branches}
\begin{subequations}
\begin{equation}
E^{\ }_{\pm}(\boldsymbol{k})=
E^{\ }_{\bar\Gamma}
+
\frac{\hbar^{2}|\boldsymbol{k}|^{2}}{2m^{*}}
\pm 
\alpha^{\ }_{{\mathrm{RB}}}|\boldsymbol{k}|.
\label{eq: dispersion}
\end{equation} 
with the corresponding eigenspinors
\begin{equation}
\left\langle\boldsymbol{k},\pm \right|=
\left(
\begin{matrix}
e^{i(\varphi \pm\pi/2)}
&,&
1
\end{matrix}
\right)
/\sqrt{2},
\label{eq: spinor}
\end{equation}
\end{subequations}
where $\varphi=\arctan (k^{\ }_{y}/k^{\ }_{x})$ {
and the two-dimensional momentum $\boldsymbol{k}$ is measured relative to
the $\bar\Gamma$ point}.
Although $H^{\ }_{{\mathrm{RB}}}$ 
breaks the spin-rotation symmetry of $H^{\ }_{0}$, 
it preserves time-reversal symmetry. 
The mechanism of the enhanced spin splitting in the 
Bi$^{\ }_{\mathsf{x}\vphantom{1}}$Pb$^{\ }_{1-\mathsf{x}}$/Ag(111) 
surface alloy goes beyond {this} 
simple model{.}
{Nevertheless,} 
many of the fundamental properties of this system 
are {captured} by the simple nearly free electron RB (NFERB) model described above .

{We plot in Fig.~\ref{Fig1}(b)
the dispersion of the NFERB model.
The dispersion along any cut passing through the $\bar\Gamma$ point
can be assigned two distinct colors that distinguish the anti-parallel
alignments of the spin expectation values for 
the eigenspinors.
This gives two spin-split bands colored in blue and red in Fig.~\ref{Fig1}(b).}
They are offset by two opposite wave vectors of magnitude 
$k^{\ }_{0}$ when measured from $\bar{\Gamma}$.
{The Rashba energy $E_{R}=\hbar^{2} k^{2}_{0}/(2|m^{*}|)$
characterizes the strength of the RB effect.}
The spin polarization vectors 
$\boldsymbol{S}^{\ }_{\pm}(\boldsymbol{k})$, 
defined as the spin expectation values of the eigenspinors 
$|\boldsymbol{k},\pm\rangle$, 
{are parallel to}
 the basal plane of Fig.~\ref{Fig1} 
and are orthogonal to $\boldsymbol{k}$, 
as depicted by the yellow arrows in Fig.~\ref{Fig1}~(b).
{Below $E^{\ }_{\bar\Gamma}$}, 
the spin polarization rotates counterclockwise 
along the outer constant energy contour and clockwise for the inner contour. 
{Above $E^{\ }_{\bar\Gamma}$}, 
the spin polarization rotates counterclockwise along both contours.
The experimentally determined spin polarization vectors 
will be denoted by $\boldsymbol{P}=(P^{\ }_{x},P^{\ }_{y},P^{\ }_{z})$
{and will be shown to obey this simple rule.

The density of states (DOS) $\nu(E^{\ }_{\mathrm{F}})$
of the NFERB Hamiltonian
is also sensitive to the change in the geometry and topology of the 
Fermi surface upon tuning of the Fermi energy
$E^{\ }_{\mathrm{F}}$. The DOS $\nu^{\ }_{\mathrm{o, i}}(E^{\ }_{\mathrm{F}})$
of the outer ($\mathrm{o}$) and the inner ($\mathrm{i}$) Fermi contour shown in Fig.~\ref{Fig1}(c) are given by
\begin{equation}
\nu^{\ }_{\mathrm{o},\mathrm{i}}(E^{\ }_{\mathrm{F}})=
\Theta(E^{\ }_{0}-E^{\ }_{\mathrm{F}})
\nu^{\ }_{2\mathrm{D}}
\left|
1
\pm
\sqrt{
\frac{
E^{\ }_{0}
-
E^{\ }_{\bar{\Gamma}}
     }
     {
E^{\ }_{0}
-
E^{\ }_{\mathrm{F}}
     } 
     }
\right|,
\label{eq: i and o DOS}
\end{equation}
whereby $\Theta$ is the Heaviside function and
$
\nu^{\ }_{2\mathrm{D}}=
|m^{*}|
/
(2\pi\hbar^{2})$.
The $+$ refers to the outer Fermi contour, the $-$ to the inner one.
The sum $\nu^{\ }_{\mathrm{o}}(E^{\ }_{\mathrm{F}})+\nu^{\ }_{\mathrm{i}}(E^{\ }_{\mathrm{F}})$ reduces to the constant DOS $2\nu^{\ }_{2\mathrm{D}}$ 
of a spin degenerate
two-dimensional hole gas with parabolic dispersion when 
$E^{\ }_{\bar{\Gamma}}>E^{\ }_{\mathrm{F}}$, 
has a singular derivative when 
$E^{\ }_{\mathrm{F}}=E^{\ }_{\bar{\Gamma}}$,
while it displays the one-dimensional Van Hove singularity
$\nu(E^{\ }_{\mathrm{F}})\sim(E^{\ }_{0}-E^{\ }_{\mathrm{F}})^{-1/2}$
in the limit $E^{\ }_{\mathrm{F}}\to E^{\ }_{0}$.}

The Bi$^{\ }_{\mathsf{x}\vphantom{1}}$Pb$^{\ }_{1-\mathsf{x}}$/Ag(111) 
sample preparation was carried out 
$in\ situ$ under ultrahigh vacuum conditions with a base pressure better than 
$2\times 10^{-10}$ mbar. The Ag(111) crystal was cleaned by multiple cycles of 
Ar$^{+}$ sputtering and annealing. Bi and Pb were simultaneously deposited 
from a calibrated evaporator at a pressure below $4\times 10^{-10}$ mbar, 
with the total amount corresponding to 1/3 of a mono-layer. 
The sample quality was affirmed by low{-}energy electron 
diffraction, which showed sharp 
$(\sqrt{3}\times\sqrt{3})R30^{\circ}$ 
spots and no further superstructure, 
and angle-resolved photoemission spectroscopy (ARPES), 
which showed a continuous tuning of the band structure and no superposition of the Bi/Ag(111) and the Pb/Ag(111) band structures. These are both strong indications that, 
although the surface is well ordered, Bi and Pb are randomly substituted.

The experiments were performed at room temperature at 
the Surface and Interface Spectroscopy beamline at the Swiss Light Source of 
the Paul Scherrer Institute using the COPHEE spectrometer \cite{moritz}. 
The data were obtained using horizontally polarized light with a photon energy 
of 24 \nolinebreak eV. Because of the inherently low efficiency of 
Mott detectors the energy and angular resolution were sacrificed 
in the spin-resolved measurements up to 80 meV and 1.5 degree, respectively. 
The coordinate system for the measurements is such that a momentum distribution
curve (MDC) is taken along the $k^{\ }_{x}$-axis. This means that a spin 
polarization vector {$\boldsymbol{P}$
is expected to point in the $\pm y$-direction 
if the NFERB model
holds qualitatively.}

A detailed description of the band structure of Bi/Ag(111) and Pb/Ag(111) 
can be found in {Refs}.~\cite{ast, pacile, bihlmayer, me}. 
{There are} two Kramer's pairs 
{K1 and K2} 
of bands that are {qualitatively} drawn in 
{Fig.~\ref{Fig1}(a)}.
The inner one (K1) is mostly of $sp^{\ }_{z}$ symmetry. 
The outer one (K2) is mostly of $p^{\ }_{x,y}$ symmetry. 
{For Pb/Ag(111) ($\mathsf{x}=0$), 
both K1 and K2 are only  partially occupied.
For Bi/Ag(111) ($\mathsf{x}=1)$, K1 is fully occupied, 
while K2 is only partially occupied}.
{Irrespective of $\mathsf{x}=0$ or $\mathsf{x}=1$, 
the spin polarization vectors for K1 are nearly parallel 
to the surface plane and are approximately perpendicular 
to the momenta, in agreement with the NFERB
model. In contrast, the spin polarization vectors for
K2 feature significant  out-of-plane components depending on the crystallographic direction. 
This is a consequence of the stronger coupling to in-plane potential gradients \cite{me, ast}. 
{We will only consider K1 from now on}.

\begin{figure}[t]
\begin{center}
\includegraphics[width=0.48\textwidth]{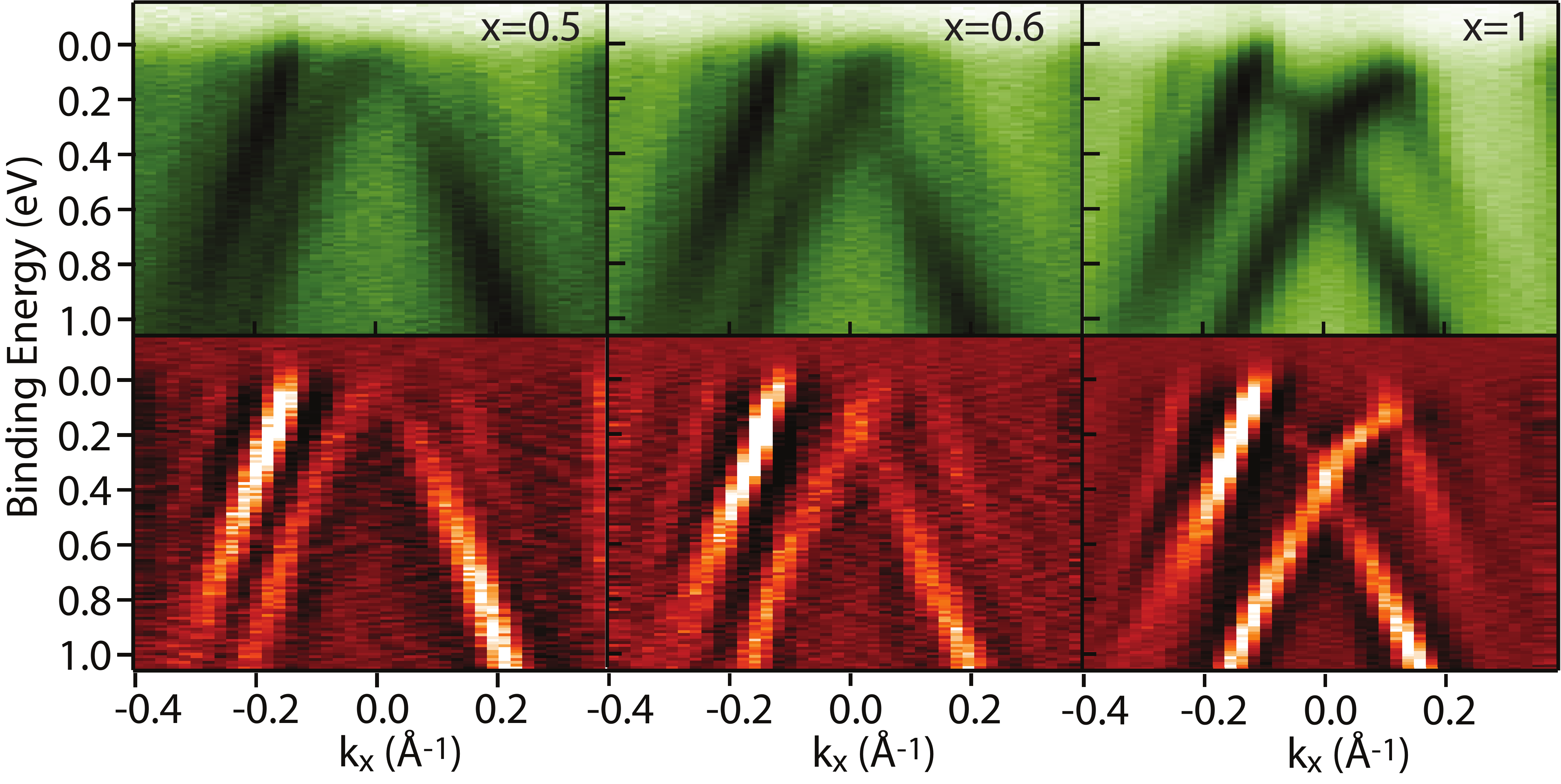}
\caption{{(color online) 
Upper graphs: Experimental band structure of 
Bi$^{\ }_{\mathsf{x}}$Pb$^{\ }_{1-\mathsf{x}}$/Ag(111) 
for $\mathsf{x}=(0.5), (0.6)$ and $(1)$ 
(from left to right) along the $\bar{\Gamma}\bar{K}$ direction, 
where dark corresponds to a higher photoemission intensity. 
Lower graphs: Second derivative data to enhance the contrast.
        }}
\label{Fig2}
\end{center}
\end{figure}

Fig.~\ref{Fig2} shows the experimental band structure of the 
Bi$^{\ }_{\mathsf{x}\vphantom{1}}$Pb$^{\ }_{1-\mathsf{x}}$/Ag(111) 
surface alloys along $\bar{\Gamma}\bar{K}$ for 
$\mathsf{x}=(0.5), (0.6)$ and $(1)$ 
measured with (spin integrated) ARPES. 
Second derivative data are also shown to enhance the contrast. 
The band K1 is fully occupied for $\mathsf{x}=1$ 
and{,} as $\mathsf{x}$ is decreased, 
the Fermi level shifts down with respect to 
{the bands so that K1 gets depopulated, 
and} the spin-splitting decreases.
For $\mathsf{x}=0.6$, 
the Fermi level $E^{\ }_{F}$ lies between the band apex 
and the crossing point ($E^{\ }_{0}>E^{\ }_{F}>E^{\ }_{\bar\Gamma}$).
{U}nconventional FSST are {then expected
according to the NFERB model. 
At $\mathsf{x}=0.5$, the Fermi level lies approximately at the crossing point of $K1$, where
the DOS of the inner Fermi contour vanishes and
a topological transition of the Fermi surface occurs according to the NFERB model. 
Note that our  calibration of $\mathsf{x}$ is slightly different from that given 
in Ref.~\onlinecite{astmixed}. {However,} this does not affect the conclusions of this work.

\begin{figure}[htb]
\begin{center}
\includegraphics[width=0.48\textwidth]{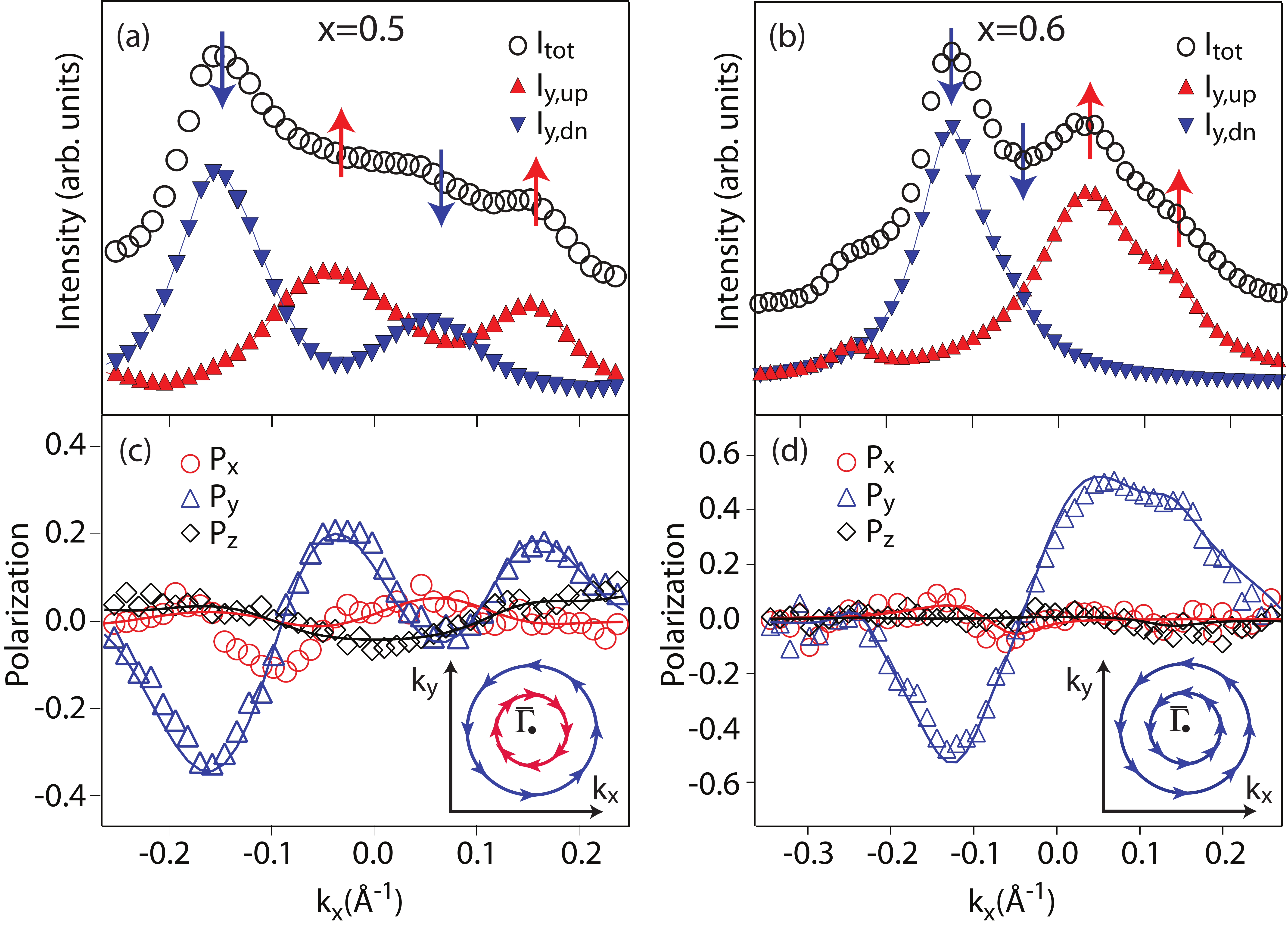}
\caption{{(color online) 
Spin resolved ARPES data of 
Bi$^{\ }_{\mathsf{x}}$Pb$^{\ }_{1-\mathsf{x}}$/Ag(111) 
for $\mathsf{x}=0.5$ (left) 
and $\mathsf{x}=0.6$ (right). 
(a) and (b) Total spin integrated intensity (circles) 
and spin-resolved intensity curves projected on the $y$-axis 
of a MDC along $\bar{\Gamma}\bar{K}$. 
(c) and (d) are the corresponding measured (symbols) and fitted (solid lines) 
spin polarization data. (Insets) Schematically drawn FSST. 
For $\mathsf{x}=0.6$, 
both bands of K1 crossing $E^{\ }_{F}$ between $\bar{\Gamma}$ 
and the SBZ boundary have parallel spin polarization vectors, 
while for $\mathsf{x}=0.5$, 
the spin polarization vectors are anti-parallel.
        }}
\label{Fig3}
\end{center}
\end{figure}

{We show in Fig.~\ref{Fig3}} 
the experimental spin-resolved MDCs for $\mathsf{x}=0.5$ 
(left {column}) and $\mathsf{x}=0.6$ (right {column}) 
providing us with the FSST for $E^{\ }_{\bar\Gamma}>E^{\ }_{F}$ and 
$E^{\ }_{0}>E^{\ }_{F}>E^{\ }_{\bar\Gamma}$, respectively. 
The extraction of the spin polarization vectors 
{$\boldsymbol{P}$} was done by applying 
a two-step fitting routine \cite{me} 
on the data of Fig.~\ref{Fig3}. For both compositions{,} 
we find that the surface states {K1} remain fully spin polarized 
with spin polarization vectors similar 
to those of the surface states of Bi/Ag(111) or Pb/Ag(111) found in \cite{me}. 
The spin polarization vectors
lie mainly in the surface plane perpendicular 
to $\boldsymbol{k}$ and both the out-of-plane and radial spin polarization 
components are comparatively small. This finding is corroborated by several 
similar measurements in different crystallographic directions 
and at different binding energies. 
We thus {c}onclude that the spin polarization 
of the surface state{s} K1 is robust against the mixing of Bi and Pb.

For $\mathsf{x}=0.5$, 
the measurement is performed slightly below the crossing point 
of K1{. W}e observe the conventional situation, 
{i.e.,} a straight cut from 
$\bar{\Gamma}$ to the {surface Brillouin zone (SBZ)} 
boundary crosses two bands with opposite spin polarization vectors. 
This can be seen in the spin-resolved spectra of Fig.~\ref{Fig3}(a), 
which are obtained from the fits of the corresponding spin polarization 
data shown in Fig.~\ref{Fig3}(c). 
The spin polarization vectors of the bands are opposite for all 
adjacent bands. 
The {corresponding qualitative} FSST 
are drawn in the inset of Fig.~\ref{Fig3}(c).

For $\mathsf{x}=0.6$, an unconventional FSST is observed. 
Fitting the spin polarization data of Fig.~\ref{Fig3}(d) 
clearly shows that, for positive and negative $k^{\ }_{x}$, 
both bands crossing the Fermi energy have {nearly parallel 
spin polarization vectors. 
The corresponding spin-resolved spectra are displayed in Fig.~\ref{Fig3}(b). 
Due to strong transition matrix element effects, 
the inner band on the left side of normal emission is only visible 
as a weak shoulder of the $I^{\ }_{y,dn}$ curve. 
When $E^{\ }_{0}>E^{\ }_{F}>E^{\ }_{\bar\Gamma}$, 
the FSST {match qualitatively those shown} 
in the inset of Fig.~\ref{Fig3}(d). 
{A}
cut from $\bar\Gamma$ {to} 
the SBZ boundary crosses two bands with parallel 
spin polarization vectors. 

\begin{figure}[htb]
\begin{center}
\includegraphics[width=0.48\textwidth]{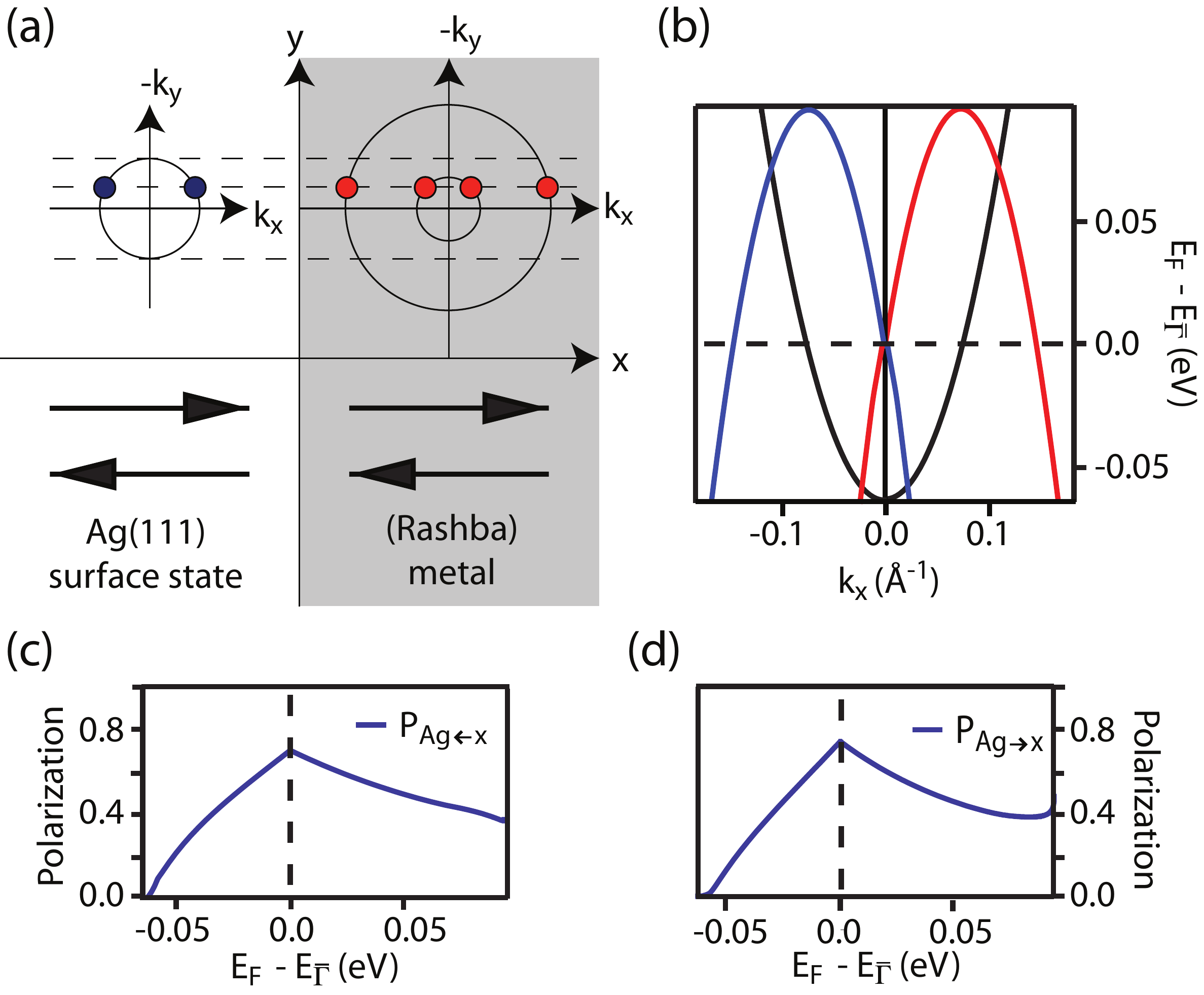}
\caption{{(color online) 
(a) Incoming and outgoing {surface} 
plane waves from the interface at $x=0$  
between a spin isotropic ($x<0$) and a {RB} ($x>0$) 
{metal}. 
(b) Dispersion{s of the Ag(111) and RB surface states}.
(c) {
$P_{\mathrm{Ag}\leftarrow\mathsf{x}}$
as a function of
$E^{\ }_{\mathrm{F}}-E^{\ }_{\bar{\Gamma}}$
as defined in the text.
(d)
$P_{\mathrm{Ag}\rightarrow\mathsf{x}}$
as a function of
$E^{\ }_{\mathrm{F}}-E^{\ }_{\bar{\Gamma}}$
as defined in the text.}
        }}
\label{Fig4}
\end{center}
\end{figure}

{
We have thus established that varying $\mathsf{x}$ 
between 0.5 and 0.6 induces a topological transition in the shape of the
Fermi surface of {K1} surface states with an impact on their spin texture
and on their DOS that is qualitatively captured by
the NFERB model. Intuitively, one could expect a spin filtering effect due to unconventional FSST,
since states with parallel $\boldsymbol{k}$-vectors posses identical spin
polarization vectors. However, it is the group velocity which determines
electronic transport and this remains the same for anti-parallel spin directions.
We will now argue that the 
transport of spins across an ideal one-dimensional boundary
separating a spin-degenerate
two-dimensional electron gas from a RB hole gas is sensitive
to this topological transition.}

{
In principle, a two-dimensional scattering geometry, 
as depicted in Fig.~\ref{Fig4}(a), 
could be realized by the deposition of 
Bi$^{\ }_{\mathsf{x}\vphantom{1}}$Pb$^{\ }_{1-\mathsf{x}}$ 
on Ag(111) through a shadow mask. We denote with
$x$ and $y$ the coordinates of the two-dimensional
Ag(111) surface. A spin-degenerate electron gas with an effective mass
corresponding to that of Ag(111) surface states meets 
the states from the K1 band of 
Bi$^{\ }_{\mathsf{x}\vphantom{1}}$Pb$^{\ }_{1-\mathsf{x}}$/Ag(111)
at the ideal one-dimensional boundary $x=0$. 
We imagine driving a small charge current through 
the boundary by applying an infinitesimal voltage difference
across the interface. 
The polarity of this applied voltage defines whether the charge current 
is from the left to the right, i.e., from the Ag(111) to the RB side, 
or from the right to the left, i.e., from the RB to the Ag(111) side.
In the Drude limit, the charge current can be calculated
from the reflection coefficients $R^{\ }_{\sigma}$ for an incoming
surface state of energy $E^{\ }_{\mathrm{F}}$ 
with spin quantum number $\sigma$ along some quantization axis,
which is here chosen to be the $y$-axis.}

We denote the spin current by 
$P_{\mathrm{Ag}\rightarrow\mathsf{x}}$
when the current is from the Ag(111) to the RB side or by
$P_{\mathrm{Ag}\leftarrow\mathsf{x}}$ otherwise.
To quantify the transport of spin across the boundary, 
we divide, \textit{on the Ag(111) side}, 
the spin current normal to the boundary 
(the difference between the spin up and spin 
down current) by the particle current normal to the boundary, i.e.
$P_{\mathrm{Ag}\leftrightarrow\mathsf{x}}=(j_{up}-j_{dn})/j_{tot}$.
We use the parameters 
$m^{*}_{\mathrm{Ag}}/m^{\ }_{e}=0.397$,
$E^{\ }_{\bar{\Gamma},\mathrm{Ag}}=-63\, meV$
on the Ag(111) side \cite{reinert} and
$m^{*}_{\mathsf{x}}/m^{\ }_{e}=-0.25$,
$E^{\ }_{0,\mathsf{x}}=94\,meV$,
$E^{\ }_{\bar{\Gamma},\mathsf{x}}=0$,
on the RB side \cite{astmixed}.
The Ag(111) and RB dispersions are shown in 
Fig.~\ref{Fig4}(b). 

We plot in Fig.~\ref{Fig4}(c)
$P_{\mathrm{Ag}\leftarrow\mathsf{x}}$,
and in Fig.~\ref{Fig4}(d) 
$P_{\mathrm{Ag}\rightarrow\mathsf{x}}$
for different band fillings as described by the value of $E^{\ }_{\mathrm{F}}$
(see Ref~\cite{Srinsongmuang08} for computational details).
In the absence of RB coupling, the spin current across
the boundary vanishes. The breaking of 
the spin{-}rotation symmetry
by the RB coupling induces a spin
current on the Ag(111) side
in Figs.~\ref{Fig4}(c) and \ref{Fig4}(d).
This induced spin current is strongly enhanced by
the onset of an unconventional FSST when the Fermi level triggers a
topological transition of the RB Fermi surface and $\nu_i$ vanishes.
Thus, the RB metal acts as a spin injector or a spin
acceptor depending on the polarity of the applied voltage difference across
the boundary.
Finally, even for non-ideal systems, such spin currents might lead to local
spin accumulation that could be detected with magnetic STM.

To conclude, we have shown that substitutional 
alloying does not alter the spin polarization vectors of the mixed 
Bi$^{\ }_{\mathsf{x}\vphantom{1}}$Pb$^{\ }_{1-x}$/Ag(111) 
surface alloys. 
Furthermore, unconventional Fermi surface spin textures were realized 
through an adequate choice of the composition and were measured. 
Systems with strong {RB} type spin{-}orbit 
splitting and 
$E_{F}\approx E_{\bar{\Gamma}}$ are suggested to function as a spin filter. 
One could also envisage using materials with similar properties 
as spin injectors for a ``classical'' {RB} system. 
This could reduce the problems encountered at interfaces to ferromagnets. 

\begin{acknowledgements}
Fruitful discussions with M. Grioni and G. Bihlmayer are gratefully acknowledged. 
We thank C. Hess, F. Dubi, and M. Kl\"ockner for technical support. 
The measurements have been performed at the Swiss Light Source, 
Paul Scherrer Institut, Villigen, Switzerland. This work is supported by the
Swiss National Foundation.
\end{acknowledgements}


\end{document}